# Post-Disturbance Dynamic Frequency Features Prediction Based on Convolutional Neural Network

Jintian Lin, Yichao Zhang, Xiaoru Wang*, *Senior Member, IEEE*, and Qingyue Chen

*Abstract*—The significant imbalance between power generation and load caused by severe disturbance may make the power system unable to maintain a steady frequency. If the post-disturbance dynamic frequency features can be predicted and emergency controls are appropriately taken, the risk of frequency instability will be greatly reduced. In this paper, a predictive algorithm for post-disturbance dynamic frequency features is proposed based on convolutional neural network (CNN) . The operation data before and immediately after disturbance is used to construct the input tensor data of CNN, with the dynamic frequency features of the power system after the disturbance as the output. The operation data of the power system such as generators unbalanced power has spatial distribution characteristics. The electrical distance is presented to describe the spatial correlation of power system nodes, and the t-SNE dimensionality reduction algorithm is used to map the high-dimensional distance information of nodes to the 2-D plane, thereby constructing the CNN input tensor to reflect spatial distribution of nodes operation data on 2-D plane. The CNN with deep network structure and local connectivity characteristics is adopted and the network parameters are trained by utilizing the backpropagation-gradient descent algorithm. The case study results on an improved IEEE 39-node system and an actual power grid in USA shows that the proposed method can predict the lowest frequency of power system after the disturbance accurately and quickly.

## I. Introduction

With the construction of the ultra-high-voltage AC and DC transmission lines and the integration of large-scale renewable energy into the power grid, the frequency dynamic behavior of the power system is becoming more complicated, which brings new challenges for power systems frequency stability. On the one hand, the tripping of high-capacity generators and large-scale renewable energy units, the damage of transmission corridor and the system splitting may cause serious imbalance of active power[1-2]. On the other hand, The decrease of system inertia, the incompatibility of operation and control strategies, and the increase of random fluctuation intensity caused by renewable energy have also brought serious challenges to power system frequency analysis and control[3], which makes frequency stability problem becoming increasingly prominent. System dynamic frequency refers to the process of frequency variation with time caused by unbalanced power[4]. After the power disturbance of system such as machine tripping, the frequency of the system will first drop to a minimum frequency, and then the frequency is restored to a quasi-steady state value by the primary frequency control. The dynamic frequency features of the power system mainly include the lowest frequency and the steady-state frequency after the disturbance. A low frequency nadir can lead to the disconnection of generators, load shedding intervention, and is therefore dangerous for system stability. Therefore, fast and accurate prediction of power system dynamic frequency features in a very short time after the disturbance occurred and the system is far from unstable is of great significance for implementing corresponding emergency frequency control strategy such as automatic load shedding and emergency DC power support, ensuring system frequency stability and preventing system frequency from collapse.

At present, the power system dynamic frequency analysis is generally based on time domain simulation[5]. Constrained by the calculation speed and the modeling accuracy, it cannot meet the requirements of large power grids online prediction. In addition, some methods for fast dynamic frequency prediction, such as equivalent model method, linearization method are proposed[6-8], which simplify the system model and obtains very fast calculation speed. However, due to a large amount of equivalent, these methods have relatively large prediction error inevitably. With the increase of power grid complexity and the decrease of system inertia caused by renewable energy integration, higher requirement for on-line frequency prediction is put forward. The current methods based on power system physical model are difficult to balance both accuracy

Manuscript received June 3, 2019. This work was supported in part by the laboratory foundation project of China Electric Power Research Institute under the project of Frequency Stability Evaluation and Control of New Energy Integrated Power System Based on Machine Learning under Grant FX83-18-002.

Jintian Lin is with Electric Power Research Institute, State Grid Zhejiang Electric Power Co.,Ltd., Hangzhou, CO 310000 China ( e-mail: jtlin@my.swjtu.edu.cn).

Yichao Zhang is with School of Electrical Engineering, Southwest Jiaotong University, Chengdu, CO 610000 China ( e-mail: yczhang_swjtu@163.com).

Xiaoru Wang is with School of Electrical Engineering, Southwest Jiaotong University, Chengdu, CO 610000 China(e-mail: xrwang@home.swjtu.edu.cn).

Qingyue Chen is with School of Electrical Engineering, Southwest Jiaotong University, Chengdu, CO 610000 China(e-mail: chenqingyue1995@163.com).

and computational efficiency.

The continuous improvement of the intelligent level of the power system and the deep integration of the power system with the Internet and the communication network provides an information foundation for the data-driven online power system frequency prediction method[9]. Taking the system operating state closely related to the disturbance degree as input, the data-driven model is used to construct the relationship from the input to the output frequency features, so that the influence of the disturbance is taken into account in the model and the fast prediction is realized. Some data-based shallow machine learning methods such as support vector machines[10], artificial neural networks[11] and decision tree[12] has been applied in power system frequency prediction. However, due to the limited feature extraction ability of shallow machine learning methods, its ability to solve complex regression and classification problems is constrained, and they cannot effectively mins and utilize and spatial-temporal correlation features existing in the power system operation data.

In recent years, deep learning[13] has subverted the dominance of shallow learning methods in many fields due to its excellent feature extraction, classification and prediction ability. It is widely used in many frontier fields such as computer vision[14], natural language processing[15] and speech recognition[16]. It also provides a new way of thinking for power system frequency analysis. Convolutional neural network (CNN) is a representative deep learning framework, which can build a deeper machine learning network with multiple hidden layer, and the training algorithms that adapt to the deep network is used to train the massive data samples to extract the spatial features of the power system operation data, thereby further improving the accuracy of power system frequency prediction. Besides, CNN extracts the feature of data by local receptive field, which is especially suitable to process data with spatial correlation features. Combined with the characteristics of power system operation data, the idea of applying CNN for frequency prediction emerges.

In this paper, a post-disturbance frequency features prediction method based on convolutional neural network is proposed. The method uses the electrical distance between nodes to describe the spatial distribution of power system buses. The dimensionality reduction algorithm based on t-distributed stochastic neighbor embedding is used to map the high-dimensional spatial distribution of nodes to the two-dimensional coordinates. Based on the 2-D coordinates of the power system nodes, the appropriate power system operating features are selected to construct the tensor input of convolutional neural networks. The CNN frame is used to extract the hidden abstract feature from the tensor input, thus the power system post-disturbance frequency features prediction model is constructed. The case study results of an improved New England 39-node system with wind farm integrated and an actual power grid of USA show that the CNN method significantly improves the prediction accuracy compared with the traditional shallow-structure frequency prediction method (SVR, ANN). Besides, comparing the depth learning method (MLP) with fully connected network and vector as input, the locally connected CNN with tensor data as input also has higher prediction accuracy benefits from its ability of extracting spatial correlation features. In addition, CNN maintains good accuracy under small training samples and unbalanced samples.

## II. CONVOLUTIONAL NEURAL NETWORK

Convolutional neural network is deep neural network with multiple layer and convolutional structure. The response of each layer of the CNN is excited by the local receptive field of the upper layer. CNN extracts the features of the input through the alternately connected convolutional layer and subsampling layer, and passes abstract features to the fully connected neural network for regression and classification analysis. A typical LeNet-5[17] convolutional neural network diagram is shown in Fig. 1, which includes input layer, several alternately connected convolutional and subsampling layers, fully connected layer, and output layer.

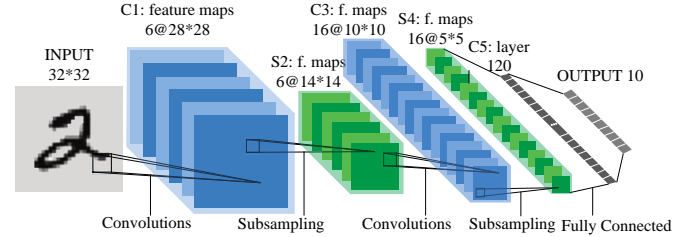

Fig .1 Structure of LeNet-5 convolutional neural network.

### A. Structure of CNN

1) Input layer: The input to the CNN is tensor data. For example, the color images are third-order tensor data obtained by superimposing the second-order tensor feature maps of three RBG channels, and the elements in the tensor matrix are the color values of the pixels. The concept of tensor is extended to the field of power system. The actual power system can be approximately regarded as distributing on a 2-D plane, and the nodes of the power system are similar to the image pixels. Using power system operation features to construct several two-order tensors, by superimposing these tensors, the third-order tensor input of the CNN can be obtained.

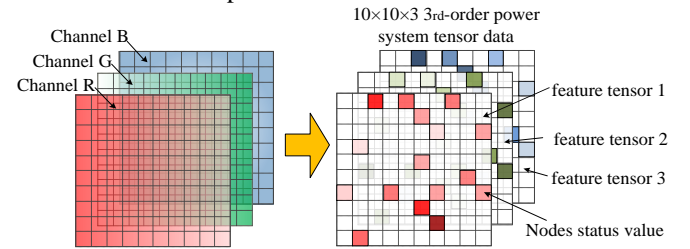

Fig.2 Analogy from RGB color images to third-order tensors of power systems.

2) Convolutional layer: The convolutional layer is sparsely connected to the local area of the previous feature map by the convolution kernel, as shown in Fig. 3. The convolution kernel is a weight matrix which slides with a certain strides on the feature map, and performs discrete convolution calculation on the date in the local area of the convolution kernel. The calculation result is transmitted to a non-linear activation function and generates the feature map of the next layer. For each feature map, the weight of the convolution kernel is constant, called the weight sharing principle, by which the amount of parameters for training can be greatly reduced.

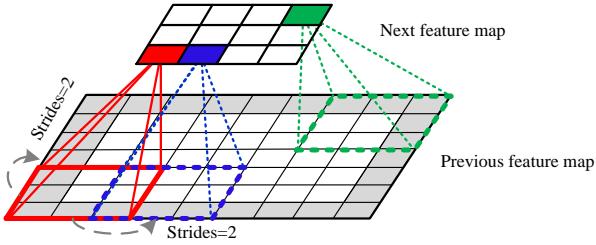

Fig. 3 Schematic diagram of local receptive field and convolution kernel.

The calculation of the convolution layer is shown in equation (1):

$$X_j^l = \sigma\left(\sum_{i \in M_i} X_i^{l-1} * K_j^l + b_j^l\right) \quad (1)$$

Where $X_j^l$ is the output of convolutional layer; $l$ is the layer number; $j$ represents the $j$-th convolution kernel, each convolution kernel corresponds to an output tensor feature map; $X_i^{l-1}$ is the $i$-th feature map of previous layer. $M_i$ is the set of input feature maps. $K_j^l$ is the weight matrix of convolution kernel; $*$ is the two-dimensional discrete convolution operator; $b_j^l$ is the bias value; $\sigma$ is the nonlinear activation function, usually adopts ReLU function.

3) Subsampling layer: The subsampling layer is usually after the convolutional layer. The outputs of the subsampling layer are also sparsely connected to upper layer feature map, and the locally connected area do not overlap. The subsampling layer performs pooling operation on the locally connected area, the pooling methods includes maximum pooling, average pooling, random pooling, as is shown in Fig. 4. The subsampling layer reduces the size of the feature map while maintains the feature scale to a certain extent, and can also avoids over-fitting problem and reduces computational expense. The calculation of the subsampling layer is shown in equation (2).

$$X_j^l = \sigma\left(\beta_j^l down\left(X_j^{l-1}\right) + b_j^l\right) \quad (2)$$

Where, $\beta_j^l$ is the weight of subsampling area; $b_j^l$ is the bias of subsampling layer; $down(\ )$ is the subsampling function.

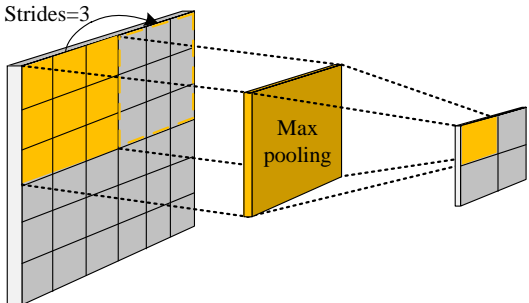

Fig. 4 Schematic diagram of the subsampling layer and pooling operation

4) Fully connected layer: After several alternately connected convolution- subsampling layer, the resulting feature map is input into the fully connected layer. In the fully connected layer, all abstract features in tensor maps are stretched and converted into vector as inputs to a fully connected neural network to obtain the final classification or regression results. The calculation of the fully connected layer is as follows:

$$X^l = \sigma\left(W^l X^{l-1} + b^l\right) \quad (3)$$

Where, $W^l$ is the weight of fully the connected layer; $b^l$ is the bias of the fully connected layer.

*B. Training of CNN*

The CNN is trained by back propagation and gradient descent algorithm. Using the loss function to calculate the error between the forward network output and the expected value. Taking the mean square error as example, there is the loss function:

$$E(K,\beta,W,b;x,y) = \frac{1}{2}\left\|h_{K,\beta,W,b}(x) - y\right\|^2 \quad (4)$$

Where, $E(K,\beta,W,b;x,y)$ is the prediction error corresponding to training sample $(x,y)$; $h_{K,\beta,W,b}(x)$ is the network feedforward calculation output corresponding to input $x$; $y$ is the expected output value.

The back propagation algorithm is used to transmit the error back to the parameters layer by layer, thus obtaining the partial derivative of each parameter to total error. The gradient descent algorithm is used to update the parameters according to a certain learning rate. By the iterative computation of backward propagation-gradient descent to minimize the error of the feedforward network and complete the training of CNN.

## III. CNN BASED POST-DISTURBANCE FREQUENCY FEATURES PREDICTION MODEL

The input of the CNN is tensor data. However, the input feature to traditional machine learning model for power system analysis is usually the data in vector form, which cannot reflect the spatial correlation of power system buses status, nor is it suitable as the input to CNN. Therefore, the power system operation data needs to be constructed into tensor form. Then the CNN is used to extract the abstract features of input data and construct mapping from the operation data input in tensor form to the dynamic frequency features after disturbance.

*A. Spatial Distribution of Nodes Based on Electrical Distance*

The propagation of electromechanical disturbances in the power system causes the state value of the system nodes to exhibit significant spatial distribution characteristics. The relationship and interaction between power system nodes are determined by the electrical distance of nodes[18]. The smaller the electrical distance, the closer the relationship between the nodes status, and the more obvious the mutual influence. The electrical distance also determines the distribution of the unbalanced power in each generator under disturbance, which is closely related to the dynamic frequency of the power system[19]. The electrical distance can be used to describe the position relations of the nodes, which is defined as the connection impedance between nodes. According to the superposition principle[20], the electrical distances between the nodes can be obtained by equation (5).

$$D_{ij} = \left| (Z_{ii} - Z_{ij}) - (Z_{ij} - Z_{jj}) \right| \tag{5}$$

Where, $Z_{ij}$ is the element of the node impedance matrix; $D_{ij}$ is the electrical distance between the nodes $i$ and $j$; For a power system with $n$ nodes, $D_{ij}$ is an n-order square symmetric matrix, the n-dimensional vector of each row of the matrix represents the electrical distance of one node to other nodes, which forms a node distribution in n-dimensional space.

*B. Dimensionality Reduction of Nodes Spatial Distribution Based on T-SNE*

The nodes spatial distribution exists in n-dimensional space is inconvenient for processing and visual display, and it is also difficult to construct the tensor input of the CNN through it. The actual power system nodes can be approximated as being distributed on a two-dimensional plane. Therefore, the idea of dimensionality reduction is naturally introduced here. As shown in equation (6). The nodes in high-dimensional space $D_{ij}$ is mapped to the 2-D plane $Y$ by dimensionality reduction, while retaining the original electrical distance relationship between nodes as much as possible.

$$D_{ij} = \begin{bmatrix} x_{11} & x_{12} & x_{13} & \cdots & x_{1n} \\ x_{21} & x_{22} & x_{23} & \cdots & x_{2n} \\ x_{31} & x_{32} & x_{33} & \cdots & x_{3n} \\ \vdots & \vdots & \vdots & & \vdots \\ x_{n1} & x_{n2} & x_{n3} & \cdots & x_{nn} \end{bmatrix} \Rightarrow Y = \begin{bmatrix} y_{11} & y_{12} \\ y_{21} & y_{22} \\ y_{31} & y_{32} \\ \vdots & \vdots \\ y_{n1} & y_{n2} \end{bmatrix} \tag{6}$$

t-distributed stochastic neighbor embedding (t-SNE)[21] is a probability-based nonlinear dimensionality reduction algorithm, which is improved by stochastic neighbor embedding (SNE). It is well-suited for embedding high-dimensional data for visualization in a low-dimensional space of two or three dimensions, and it's also one of the most effective dimension reduction algorithms at present.

The basic principle of the SNE is to convert the Euclidean distance between the respective nodes coordinates in the high-dimensional and the low-dimensional space into the conditional probability, and try to minimize the deviation of the two conditional probability, so as to maintain the distance relationship between nodes before and after dimensionality reduction.

Redefining the electrical distance matrix $D_{ij}$ as a high-dimensional spatial data $X$.

The conditional probability $p_{j|i}$ between each row vector in the high-dimensional spatial data $X$ is defined as:

$$p_{j|i} = \frac{\exp\left(-\|x_i - x_j\|^2 / 2\sigma_i^2\right)}{\sum_{k \neq i} \exp\left(-\|x_i - x_k\|^2 / 2\sigma_i^2\right)} \tag{7}$$

Where, $x_i$ and $x_j$ is the node in the high-dimensional space $X$, which is represented by the n-dimensional vector of the $i$-th and the $j$-th row of the high-dimensional space $X$; $P_{j|i}$ is the conditional probability that the node $x_j$ appears near the node $x_i$; $\sigma_i$ is the variance of the Gaussian function centered on the node $x_i$; $\|x_i - x_j\|$ is the Euclidean distance between the node $x_i$ and $x_j$.

Assuming that the high-dimensional spatial data $X$ of the power system nodes is mapped to $Y$ in the low dimension, then there is also the conditional probability $q_{j|i}$ in the low-dimensional space $Y$:

$$q_{j|i} = \frac{\exp\left(-\|y_i - y_j\|^2 / 2\sigma_i^2\right)}{\sum_{k \neq i} \exp\left(-\|y_i - y_k\|^2 / 2\sigma_i^2\right)} \tag{8}$$

Calculating the Kullback-Leibler divergence between the two probability distributions to measure the deviation between the two probability distributions and set it as the objective function $C$.

$$C = \sum_i KL(P_i \| Q_i) = \sum_i \sum_j p_{j|i} \log \frac{p_{j|i}}{q_{j|i}} \tag{9}$$

Minimize the objective function by the gradient descent algorithm, thereby obtaining the optimal nodes coordinates in low-dimensional space.

However, the Kullback-Leibler divergence objective function of SNE is asymmetrical, the penalty coefficient corresponding to different distances are different, which makes the SNE algorithm tend to retain the local structure in high-dimensional data and ignore the global features, while the asymmetric objective function is also difficult to optimized. Besides, the "congestion problem" caused by the different characteristics of high-dimensional and low-dimensional space makes the boundaries between different clusters in dimensionality reduction result very blurred.

The t-SNE algorithm utilizes the joint probability distribution to replace the conditional probability distribution, thereby eliminating the asymmetry of the objective function. At the same time, the t-SNE algorithm uses the Gaussian distribution to convert the Euclidean distance in the high-dimensional space $X$, and uses the t-distribution with a freedom degree of 1 to replace the Gaussian distribution in the low-dimensional space $Y$, thereby, the close nodes in the high-dimensional space will be mapped closer in low-dimensional space, and nodes far away will be mapped farther, thus solving the problem of nodes congestion.

The joint probability of high-dimensional space and low-dimensional space data is as follows:

$$p_{ij} = \frac{p_{i|j} + p_{j|i}}{2n} \tag{10}$$

$$q_{ij} = \frac{\left(1 + \|y_i - y_j\|^2\right)^{-1}}{\sum_{k \neq i} \left(1 + \|y_k - y_i\|^2\right)^{-1}} \tag{11}$$

Where, $n$ is the number of nodes.

The new objective function is derived as:

$$C = \sum_i KL(P\|Q) = \sum_i \sum_j p_{ij} \log \frac{p_{ij}}{q_{ij}} \tag{12}$$

By minimizing the new objective function, the optimal 2-D nodes coordinates $Y$ after t-SNE dimensionality reduction can be obtained by gradient descent algorithm.

$$\frac{\delta C}{\delta y_i} = 4 \sum_j (p_{ij} - q_{ij})(y_i - y_j)\left(1 + \|y_i - y_j\|^2\right)^{-1} \tag{13}$$

*C. CNN input tensor construction*

To construct the tensor input of CNN, the matrix elements corresponding to the 2-D node coordinates are marked in the tensor matrix and the elements are filled with the power system operation data of the nodes to obtain the feature map in tensor form..

Since the 2-D spatial distribution $Y$ is a set of coordinates in the interval of $[0,1]$, the index of matrix elements are integers, in order to mark the elements in the tensor matrix corresponding to nodes coordinates, choosing an appropriate integer $h$ and then the 2-D coordinates of nodes are enlarged to the interval of $[1,h]$ by linear normalization, Round the normalized nodes coordinates and the and the coordinates $Y_{\text{int}}$ of the integer form are obtained. As is shown in equation (14).

$$Y_{\text{int}} = round\left[1 + \frac{(h-1)(Y - Y_{\min})}{Y_{\max} - Y_{\min}}\right] \tag{14}$$

Construct a tensor matrix of size $h \times h$, The operation status value of power system nodes is assigned to the tensor matrix elements corresponding to the nodes coordinates $Y_{\text{int}}$, and assigns remaining matrix elements that do not correspond to the nodes to 0, thus a tensor feature map corresponding to a power system operation feature data is constructed, as is shown in equation (15).

$$M_t(Y_{\text{int},i1}, Y_{\text{int},i2}) = S_i^t, t \in T, i = 1, 2, \cdots, n \tag{15}$$

Where $Y_{\text{int},i1}$ and $Y_{\text{int},i2}$ are the abscissa and ordinate of the $i$-th power system node; $M_t(Y_{\text{int},i1}, Y_{\text{int},i2})$ is the value of tensor matrix element corresponding to the node coordinates; $t$ is the type of power system operation feature; $S_i^t$ is the operation feature value of the $i$-th power system node; $T$ is the set of power system operating data type.

Inspired by the tensor data of the RBG color map, different power system features are assigned to several second-order tensor feature maps, and these feature maps are superimposed to construct the third-order tensor input to the CNN.

$$\text{INPUT} = \{M_1, M_2, \cdots, M_t\} \tag{16}$$

*D. Input feature selection*

In order to balance the accuracy and the training efficiency of the CNN, it is necessary to select the appropriate power system operation feature that affects the dynamic frequency of the power system as the input to CNN.

During the dynamic process of the power system, the frequency of the system fluctuates around the inertia center of the system[22]. The inertial center frequency is usually used to represent the global characteristics of the power system dynamic frequency, which is defined as follows:

$$\omega_{sys} = \frac{\sum_{i=1}^n (H_i \omega_i)}{\sum_{i=1}^n H_i} \tag{17}$$

Where, $\omega_{sys}$ is the system inertia center frequency; $H_i$ is inertia time constant of the $i$-th generator in power system; $\omega_i$ is the frequency of the $i$-th generator in system, and the generators of the nodes are located at nodes.

Defining the moment before the disturbance occurs as $t_0$, and the moment after the disturbance occurs as $t_f$. In the first stage of the power system dynamic process, that is, at the moment when the disturbance occurs, the generators will assume the unbalanced power according to the respective synchronization coefficients, which is related to the initial operation state of the generators and the electrical distance to the fault point[18]. The synchronization coefficient $P_{sik}$ between the generator node and other nodes is defined as:

$$P_{sik} = V_i V_k (B_{ik} \cos \delta_{ik} - G_{ik} \sin \delta_{ik}) \tag{18}$$

Where, $V_i$, $V_k$ and $\delta_{ik}$ are the voltage amplitude and phase difference respectively; $B_{ik}$ and $G_{ik}$ are the transfer impedance.

On this basis, the electromagnetic power of each generator, the active power of each load at the moment $t_0$, and the voltage and phase angle of each node and the unbalanced power of each generator at the moment $t_f$ are selected as input feature.

In the second stage of the power system dynamic process, the generator rotor is affected by the unbalanced power and begins to gradually change the speed. The power system dynamic frequency is directly determined by the motion equation of the generator rotor, for the multi-machine power system, there is the dynamic frequency equation as follows[23]:

$$\dot{\omega}_{sys} = \frac{1}{2H_{sys}}\left(\sum_{i=1}^n P_{mi} - \sum_{i=1}^n P_{ei} - \sum_{i=1}^n D\omega_i\right) \tag{19}$$

Where, $H_{sys}$ is the sum of the generators' inertia time constants; $P_{mi}$ and $P_{ei}$ are the electromagnetic power and mechanical power of generators, $D$ is generator damping coefficient.

According to the dynamic frequency equation (19), the electromagnetic power and mechanical power of the generators at the moment $t_f$ are selected as input feature. In order to reflect the influence of the generators with different inertia to the dynamic frequency, the response of each generator to the frequency are introduced[11].

$$f_i = \frac{P_{mi} - P_{ei}}{2H_i} - \frac{\sum_{i=1}^n P_{mi} - \sum_{i=1}^n P_{ei}}{2H_{sys}} \tag{20}$$

In the third stage of the disturbance, the prime movers and governor begins to response to the frequency fluctuation, and the frequency is gradually restored by regulating the prime mover's output. The effect of the regulation is affected by the reserve capacity and the dynamic characteristics of the prime movers and governors. Traditional frequency response models based on equivalent physical model ignore the nonlinear links such as the limiter of the governor, and the influence of the reserve capacity of the generators cannot be taken into account. Therefore, the spare capacity of the generators $P_{ri}$ at the moment $t_f$ is introduced here as the input feature, which is defined as:

$$P_{ri} = P_{\max i} - P_{ei} \quad (21)$$

Where $P_{ri}$ is the reserve capacity of each generator; $P_{\max i}$ is the rated active power output of each generator.

Besides, considering the frequency and voltage-dependent characteristics of the load, the load power of each node are selected as input feature.

Based on the analysis above, the original input feature set for dynamic frequency prediction is obtained, which consists of 14 power system operation features. As shown in the table below.

TABLE I
ORIGINAL INPUT FEATURE SET FOR DYNAMIC FREQUENCY PREDICTION

| Number | Input Feature type |
|---|---|
| 1 | Electromagnetic power of each generator at $t_0$ |
| 2 | Mechanical power of each generator at $t_0$ |
| 3 | Reserve capacity of each generator at $t_0$ |
| 4 | Voltage amplitude of each node at $t_0$ |
| 5 | Voltage phase angle of each node at $t_0$ |
| 6 | Load active power at $t_0$ |
| 7 | Electromagnetic power of each generator at $t_f$ |
| 8 | Mechanical power of each generator at $t_f$ |
| 9 | Reserve capacity of each generator at $t_f$ |
| 10 | Power shortage of each generator at $t_f$ |
| 11 | Response of generators to disturbances at $t_f$ |
| 12 | Voltage amplitude of each node at $t_f$ |
| 13 | Voltage phase angle of each node at $t_f$ |
| 14 | Load active power at $t_f$ |

In order to reduce the occupancy of computer storage space and improve the training efficiency of CNN, based on the collected operation data of power system. The spearman correlation[26] coefficient is used to evaluate the correlation between each type of input feature and the dynamic frequency features after disturbance and the input features is further selected according to the correlation analysis result. The spearman correlation coefficient is defined as :

$$\rho_s = 1 - \frac{6\sum d_i^2}{N(N^2 - 1)} \quad (22)$$

Where, $\rho_s$ is the spearman correlation; $d_i$ is the difference between the rank order of input feature and frequency freture; $N$ is the number of samples.

*E. Frequency prediction modeling*

The construction process of the power system disturbance frequency prediction model based on CNN is shown in Fig .5 and the specific steps are as follows：

1) Obtaining the power system nodes admittance matrix according to the topological relationship of the nodes and branches, inverts it and get the nodes impedance matrix. Using the node impedance matrix to calculate the electrical distance between nodes by equation (5) to describe the nodes distribution in high-dimensional space.

2) The t-SNE algorithm is used to reduce the dimensionality of nodes spatial distribution and map it to 2-D plane. The nodes coordinates obtained after dimensionality reduction are normalized and rounded to get the integer coordinates between the interval $[1, h]$ according to equation (14).

3) Gathering the power system operation data during the disturbance through the measurement equipment or the offline simulation. Select the operation data of nodes in Table I to form the original input feature data in vector form, selected dynamic frequency features as output data. Perform data pre-processing such as data cleaning and normalization.

4) Constructing a matrix of size $h \times h$, assigns the selected operation features to the matrix elements corresponding to the coordinates of the nodes, and obtains the input samples of the CNN in form of third-order tensor.

5) The tensor input samples are divided into training set and testing set in proportion by random sampling, using the training set to train the parameters of the CNN.

6) Using the trained CNN model to predict the post-disturbance dynamic frequency features of the test set and evaluate the prediction accuracy.

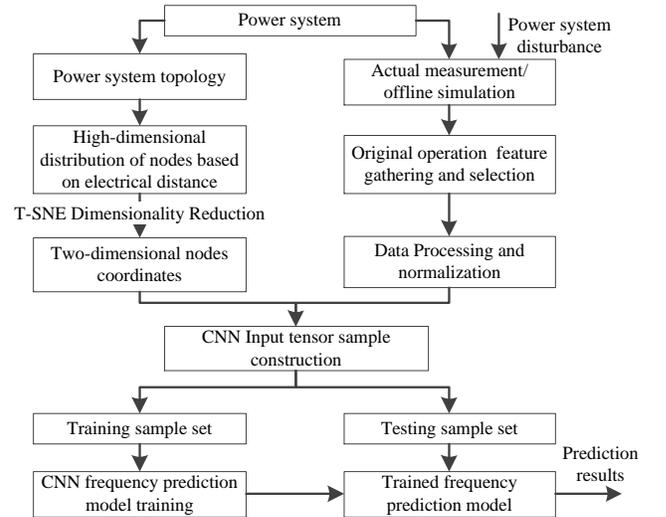

Fig .5 Construction process of the power system post-disturbance frequency prediction model based on CNN.

IV. CASE STUDY

In order to test the performance of the proposed power system dynamic frequency features prediction method based on CNN, an improved New England 39-node system with wind farm integrated and an power grid of an actual state in the United States is used for case study.

## A. Improved New England 39-node system with wind farm integrated

### 1) System introduction

The improved New England 39-node system with wind farm integrated has 11 generators and 40 nodes. Among them, the No. 30-39 nodes are conventional generators, and the No. 40 node is a wind farm with a rated capacity of 900 MW. The total active power output of the generators in the initial operation mode is 6206 MW. The wiring diagram of the system is shown is Fig. 6.

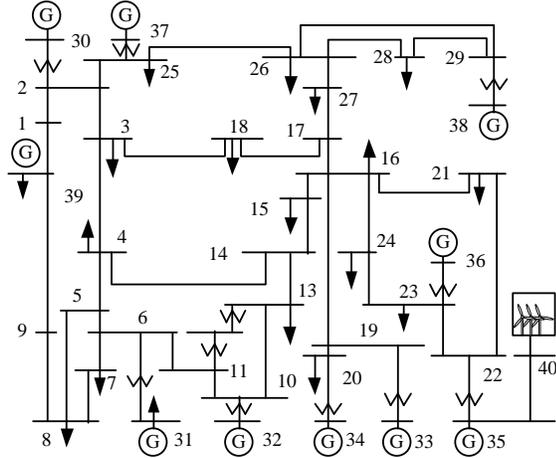

Fig .6 Wiring diagram of the improved New England 39-node test system.

### 2) Input tensor data construction

The system is set to operate at 20 load levels from 50% to 100%. Under different load levels, set the generators tripping faults and load fluctuation disturbances occur at 0 second, the system operation features in Table I are recorded as CNN input. The lowest frequency feature of the system after the disturbance are recorded as output. The simulation duration is set to 60 seconds. A total of 1400 sets of data are generated by simulation software PSS/E.

Using the spearman correlation coefficient to evaluate the correlation between each type of feature and the lowest frequency after disturbance. According to the spearman correlation analysis results, six input operation features with the highest frequency correlation coefficient are further selected as the input feature to CNN, as is shown in Table II. Each type of input feature is linearly normalized to the [0,1] interval to eliminate the effect of the feature dimension.

TABLE II
INPUT FEATURE FURTHER SELECTED BY SPEARMAN CORRELATION

| Input feature type | Spearman correlation coefficient |
| --- | --- |
| response of generators to disturbances at $t_f$ | 0.5591 |
| power shortage of each generator at $t_f$ | 0.4731 |
| voltage phase angle of each node at $t_f$ | 0.4107 |
| electromagnetic power of each generator at $t_f$ | 0.3613 |
| electromagnetic power of each generator at $t_0$ | 0.2913 |
| load active power at $t_f$ | 0.2898 |

Calculating the electrical distance between nodes to form the high-dimensional nodes spatial distribution $X$, as shown in Table III:

TABLE III
HIGH-DIMENSIONAL NODES SPATIAL DISTRIBUTION

| Node number | Electrical distance between nodes ($\Omega$) | | | | | | |
| --- | --- | --- | --- | --- | --- | --- | --- |
| | 1 | 2 | 3 | … | 38 | 39 | 40 |
| 1 | 0 | 0.0322 | 0.0387 | … | 0.1017 | 0.0216 | 0.1312 |
| 2 | 0.0322 | 0 | 0.0122 | … | 0.0724 | 0.0420 | 0.1070 |
| 3 | 0.0387 | 0.0122 | 0 | … | 0.0737 | 0.0448 | 0.0997 |
| … | … | … | … | … | … | … | … |
| 38 | 0.1017 | 0.0724 | 0.0737 | … | 0 | 0.1094 | 0.1572 |
| 39 | 0.0216 | 0.0420 | 0.0448 | … | 0.1094 | 0 | 0.1359 |
| 40 | 0.1312 | 0.1070 | 0.0997 | … | 0.1572 | 0.1359 | 0 |

The t-SNE dimensionality reduction algorithm is used to map the high-dimensional nodes spatial distribution to the 2-D space $Y$, and the 2-D coordinates in $Y$ are linearly normalized to the interval [1,100] and rounded. As shown in Table IV.

TABLE IV
THE ENLARGED 2-D NODES COORDINATES AFTER LINEAR NORMALIZATION

| Node number | The enlarged two-dimensional coordinates of nodes | |
| --- | --- | --- |
| | $Y_{int,i1}$ | $Y_{int,i2}$ |
| 1 | 54 | 84 |
| 2 | 65 | 41 |
| 3 | 67 | 14 |
| … | … | … |
| 38 | 1 | 47 |
| 39 | 51 | 100 |
| 40 | 6 | 78 |

The distribution figure of nodes in the 2-D plane is draw according to the coordinates in Table IV, as shown in Fig. 7.

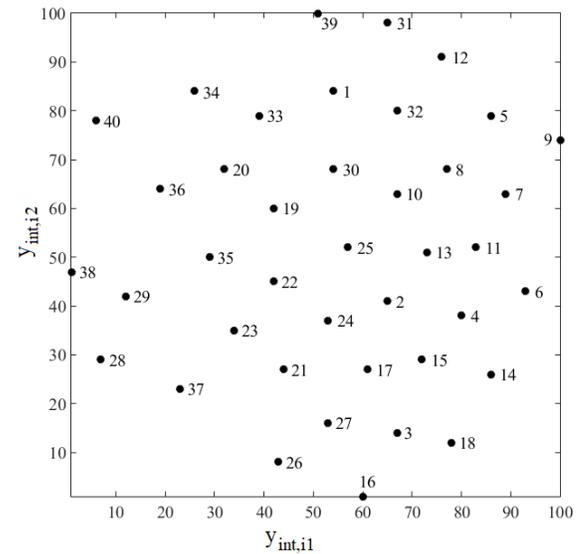

Fig. 7 Distribution figure of nodes

The effect of t-SNE dimensionality reduction can be visually reflected in Fig. 7. Taking node 15 as an example, the electrical distances of the node 15 to nodes 14, 17, and 18 are 0.0163, 0.0145, and 0.0188, respectively. Correspondingly, the node 15 is mapped closer to these nodes after dimensionality reduction. The electrical distances from node 15 to nodes 36, 37, and 38 are 0.0553, 0.0554, and 0.0813, respectively, and they are mapped further in the two-dimensional plane.

Define a matrix $M$ of size $100 \times 100$, and mark the elements $M(Y_{int,i1}, Y_{int,i2})$ in the matrix corresponding to the 2-D coordinates of the nodes. Assigning one kind of input feature of the nodes to the corresponding matrix elements, thereby obtaining a second-order tensor feature map corresponding to the feature type.

The second-order tensor feature maps corresponding to six input feature of power systems are superimposed to obtain a third-order tensor data of size $100\times100\times6$. Fig. 8 visually shows a second-order tensor data corresponding to the phase angle of each node at $t_f$ (generator tripping at No.30 node, load level 100%). The elements are given different colors according to the input operation feature values.

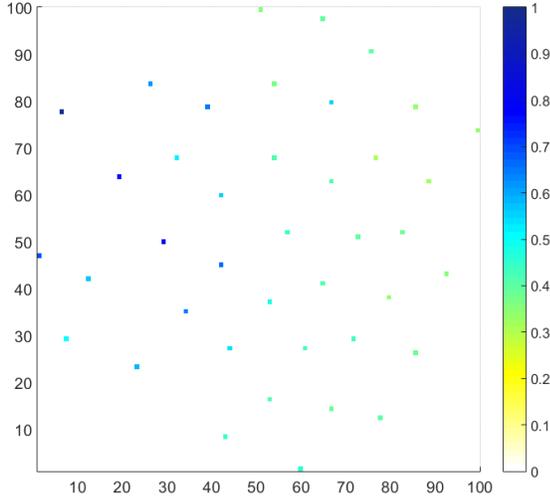

Fig. 8 Visualization of the tensor data corresponding to nodes voltage phase angle.

Reconstruct all 1400 sets of simulation data into third-order tensor data and the input sample data of the CNN is obtained.

*3) Post-disturbance frequency feature prediction*

Randomly selecting 1000 samples as training samples, and the remaining 400 are used as test samples. The CNN, multi-layer perception (MLP), support vector regression (SVR), and artificial neural networks (ANN) are used to predict the lowest frequency after power system disturbance respectively. Among them, MLP is a fully connected neural network with deep structure, ANN and SVR are shallow machine learning model.

At present, the hyperparameters of the deep learning model can only be determined by repeated test, that is, changing the hyperparameters of the network for training, and selecting the set of hyperparameters with the relatively smallest error and the best prediction result. According to a series of tests, the hidden layer of the convolutional neural network is set to 5 layers, which are convolution layer, subsampling layer-convolution layer, subsampling layer and fully connected layer. The number of convolution kernels or pooling kernels of each layer is 32-32-64-64 respectively, the number of neurons in the fully connected layer is 256. The window size of the convolution kernel is $10\times10$, the sliding strides of window is 1, the convolution kernel activation function is ReLU, the subsampling window size is $2\times2$, the window sliding strides is 2, and the subsampling method adopts the maximum pooling. The activation function of the fully connected layer is the Tanh function. The backward propagation-gradient descent algorithm is used to[24] is used to update the network parameters. The learning rate is 1e-6 and the training epoch is 1000. Referring to the structure of the CNN, the hidden layer of the MLP is set to 5 layers, and the number of neurons in each layer is set to 32-32-64-64-256 according to the kernels and neurons in each layer of CNN. Both MLP and ANN are trained by gradient descent algorithm. The support vector regression model uses the velocity modified particle Swarm optimization (MVPSO)[25] to search for the optimal penalty factor and radial basis kernel function parameters.

Using each trained machine learning model to predict the testing sample, Based on prediction results obtained by different model, the mean absolute error (MAE), mean absolute percentage error (MAPE) and root mean square error (RMSE) are used to evaluate the prediction accuracy. The above accuracy indicators are defined as follows:

$$\text{MAE} = \frac{1}{N}\sum_{i=1}^{N}\left|f(x_i)-a_i\right| \quad (23)$$

$$\text{MAPE} = \frac{1}{N}\sum_{i=1}^{N}\frac{\left|f(x_i)-a_i\right|}{\left|f(x_i)\right|} \quad (24)$$

$$\text{RMSE} = \sqrt{\frac{1}{N}\sum_{i=1}^{N}(f(x_i)-a_i)^2} \quad (25)$$

Where $N$ denotes the number of test samples, $f(x_i)$ denotes the lowest frequency predicted, $a_i$ denotes actual lowest frequency after disturbance.

Considering the randomness of the training process, repeat the training process for several times, and the average error analysis results obtained are shown in the Table V:

TABLE V
ACCURACY ANALYSIS RESULTS OF DIFFERENT PREDICTION METHOD

| Prediction method | MAE/Hz | MAPE | RMSE |
|---|---|---|---|
| CNN | 0.0018 | 3.0698e-05 | 0.0024 |
| MLP | 0.0030 | 5.0298e-05 | 0.0109 |
| SVR | 0.0041 | 6.8178e-05 | 0.0105 |
| ANN | 0.0043 | 7.1977e-05 | 0.0125 |

According to the accuracy analysis results, the proposed power system post-disturbance dynamic frequency featureslo prediction method based on CNN significantly improves the accuracy compared with the traditional shallow machine learning method (SVR, ANN). The mean absolute error of CNN compared to two above methods is reduced by 56.1%, 58.1, respectively.

Compared with MLP with deep network structure, the proposed method also has superior performance. On the one hand, the input of MLP is vector form, which erases the spatial correlation information existing in data. On the other hand, the network of MLP It is fully connected, and it is more likely to have problems such as overfitting, which affects the accuracy of the network. The tensor input with spatial features and the deep network structure with local connections of CNN greatly improvs these problems.

Since the large frequency disturbances rarely occurs in actual power system, the sample for training is usually limited. The influence of the training sample number on the accuracy of different prediction method is analyzed in this paper. The methods were trained by 50, 200, 400, 600, 800, 1000 sets of samples, and the same test set was used for prediction. Using the mean absolute error to analyze the accuracy of the prediction results. The mean absolute error under different number of training samples is shown in Table VI.

TABLE VI
THE AVERAGE ABSOLUTE ERROR OF THE TEST SAMPLE CORRESPONDING TO THE NUMBER TRAINING SAMPLES AND DIFFERENT METHODS

| Prediction method | Training sample number | | | | | |
|---|---|---|---|---|---|---|
| | 50 | 200 | 400 | 600 | 800 | 1000 |
| CNN | 0.0093 | 0.0068 | 0.0038 | 0.0027 | 0.0023 | 0.0017 |
| MLP | 0.0425 | 0.0080 | 0.0039 | 0.0032 | 0.0032 | 0.0030 |
| SVR | 0.0177 | 0.0116 | 0.0102 | 0.0079 | 0.0057 | 0.0041 |
| ANN | 0.0588 | 0.0091 | 0.0047 | 0.0040 | 0.0035 | 0.0043 |

According to the accuracy evaluation results in Table VI, the prediction error of various machine learning methods under different training samples is shown in Fig. 9.

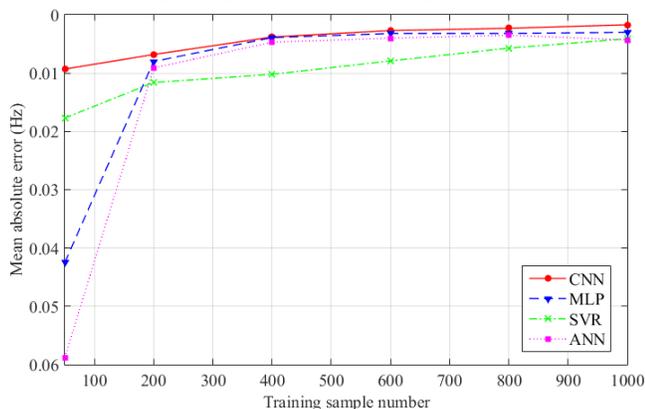

Fig. 9. Mean absolute error of different methods with varying training sample number.

According to the prediction accuracy analysis results shown in Table VI and Fig. 9, it can be seen that with the increase of the number of training samples, the prediction accuracy of various machine learning models is gradually increasing, under different sizes of training samples. The frequency prediction method based on CNN proposed in this paper has smaller prediction error than other machine learning methods. In particular, in the case of only 50 training samples, the CNN method also has high accuracy, which has significant advantages over other machine learning methods. The mean absolute error is reduced by 78.1%, 47.5%, and 84.2% compared to MLP, SVR, and ANN. Besides, MLP and ANN with fully connected networks have severe over-fitting problem in the case of small sample size.

Selecting 50 sample in the test set, as shown in Fig.10. It can be seen that the prediction results of CNN are closer to the actual values than other methods. In addition, CNN method has a fast calculation speed, which only costs 4.063 ms for single prediction. The above analysis shows the frequency prediction method based on CNN can analyze the frequency after power system disturbance quickly and accurately.

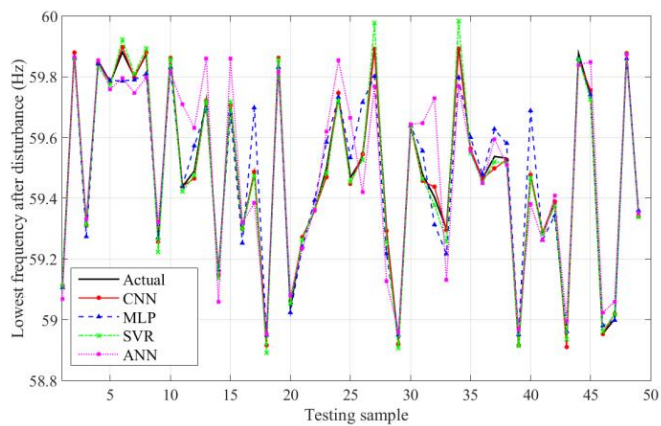

Fig. 10. Comparison of the predicted results with the actual values

### B. Actual power grid in the United States

The actual large power grid usually has strong spatial distribution feature information. In order to test the application ability of the CNN-based frequency prediction method on the actual large power grid, this section introduces the actual power system of an actual state in the United States as an example.

*1) System introduction*

The power system of this state has 500 nodes and the voltage levels of nodes are 345kV, 138kV, 13.8kV, there are 466 transmission lines, 131 transformers, and 90 generators in this power system, including thermal generators, hydropower generators and gas generators. The installed capacity of the system is 14626.74MVA. The geographic wiring diagram of this actual power system is shown in Fig. 11.

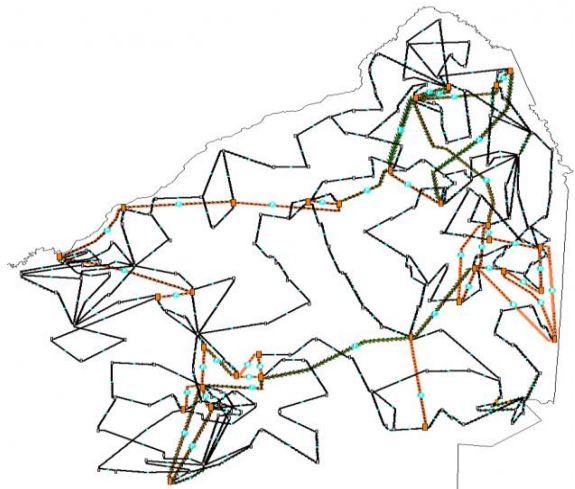

Fig. 11. Geographic wiring diagram of an actual power system in USA

*2) Analysis of spatial distribution characteristics of power system operation feature*

The operation feature of the actual large power grid disturbance appears a strong spatial distribution characteristic. Firstly, the electrical distance is used to describe the location relationship of each node in the system, and then the t-SNE dimensionality reduction algorithm is used to map the nodes to the 2-D space. The distribution of the power system nodes in the 2-D plane is as shown in Fig. 11. The spatial distribution of the nodes can be clearly seen from the figure, which can reflect the location relationship of the nodes to a certain extent.

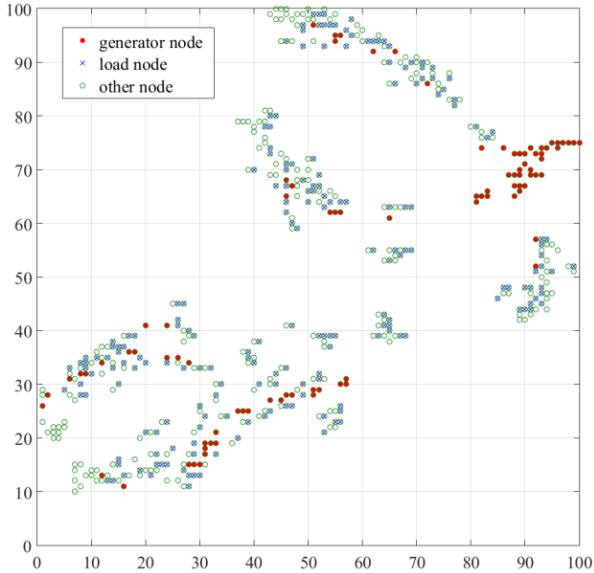

Fig. 11 distribution of the power system nodes in the 2-D plane

Using the same post-disturbance operation data generation method to obtain 1400 sets of disturbance simulation data, and then these data is reconstructed into tensor form and the tensor feature map corresponding to each type of operation data is obtained.

Using the same post-disturbance operation data generation method and input tensor construction method, the sample data of CNN samples is obtained. Fig. 12 clearly shows the operation status values of nodes at different locations on a 2-D plane under a certain disturbance, which forming several second-order tensor feature maps. They are the generator electromagnetic power, the load active power and the node voltage phase angle at the moment $t_f$

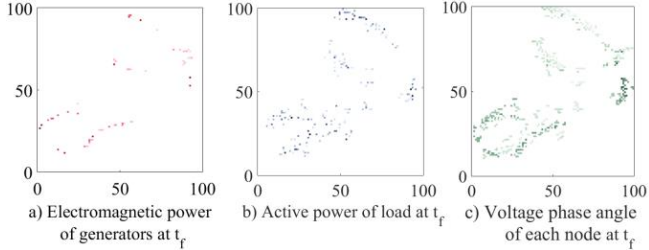

Fig.12 Power system operation status feature map

Fig. 13 shows the unbalanced power of each generator at the moment that the node 82 and node 439 occurs generator tripping fault. The matrix coordinates corresponding to node No. 82 and the node No. 439 are (46, 65) and (9, 32), respectively. It can be clearly seen from the Fig. 13 that after the fault occurs, the nodes closer to the faulty node is more affected by the fault, and tend to assume more unbalanced power. This again proves that the CNN tensor input construction method proposed in this paper can maintain the spatial correlation information of the power system node operation data.

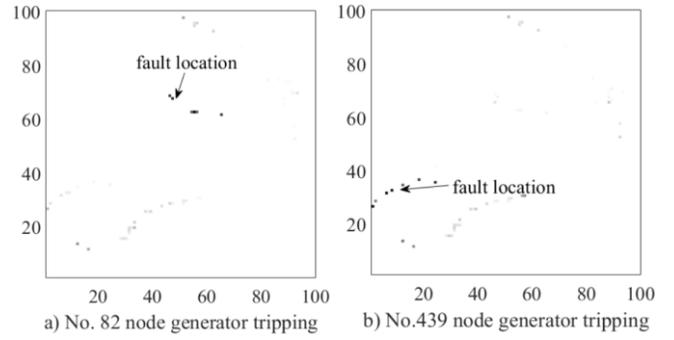

Fig.13 Feature map of the unbalanced power under the fault at different locations

### 3) Post-disturbance frequency prediction

The power system operation data in the form of third-order tensor is used as the input to the CNN, the lowest frequency after the power system disturbance is used as the output. Randomly selecting 1000 sets of samples as training samples, and the rest as testing samples. When training CNN, because this actual power system has a large number of nodes and the nodes distribution is dense, the size of the convolution kernel is appropriately reduced to 5×5 to better mines the local features of the input, other hyperparameter are the same as the previous example. Using the trained CNN model to predict the testing set, and the accuracy analysis result is shown in Table VII.

TABLE VII
ACCURACY ANALYSIS RESULTS OF DIFFERENT PREDICTION METHOD

| Prediction method | MAE/Hz | MAPE | RMSE/Hz |
| --- | --- | --- | --- |
| CNN | 0.0021 | 3.4964e-05 | 0.0030 |
| MLP | 0.0027 | 4.4204e-05 | 0.0042 |
| SVR | 0.0032 | 5.3661e-05 | 0.0037 |
| ANN | 0.0056 | 9.2802e-05 | 0.0065 |

The accuracy analysis results of the testing sample show that the power system post-disturbance frequency prediction method based on CNN also has higher accuracy when applied in the actual power system.

Fig. 14 shows the prediction results of the testing samples. As can be seen from the figure, the lowest frequency predicted by CNN model is very close to the actual simulation results. In addition, due to the large capacity of the system, most power fluctuations only cause slight frequency disturbances, and only a few power disturbances cause large frequency deviations, of which 7.75% exceeds 0.1 Hz, and 1.5% exceeds 0.2 Hz. It can be seen that in the face of such poorly balanced samples, the CNN method also has a strong generalization ability, which can accurately predict the lowest frequency fluctuation caused by large power disturbance.

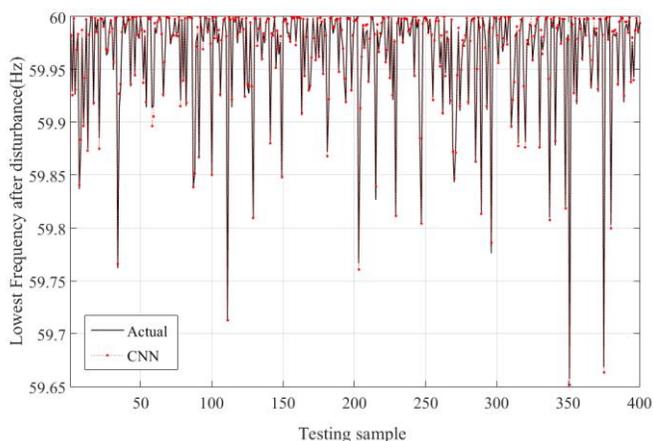

Fig. 14 Comparison of CNN prediction results and actual simulation values.

## V. Conclusion

CNN with its tensor data input and deep network structure has excellent processing ability for spatial correlated data. In this paper, a CNN based predictive method for power system post-disturbance frequency features is proposed. By constructing the power system operation data into tensor form, the spatial correlation information of the operation data is effectively retained in the CNN input data, thus laying a good foundation for the successful application of CNN to the accurate prediction of post-disturbance frequency features. The case study shows the proposed frequency prediction method significantly improves the accuracy compared to the traditional shallow machine learning method such as SVR, ANN. Compared to the deep learning MLP based method with fully connected network and vector as input, the proposed CNN based one also has higher accuracy due to its locally connected structure and tensor form input. Furthermore, the proposed method has better performance of generalization ability in the case of small samples and unbalanced samples of large power grid. The proposed method can be applied to the online frequency stability analysis of the actual power system. According to the prediction results such as maximum frequency deviation, the corresponding emergency control such as automatic load shedding can be formulated to prevent the power system from frequency instability.

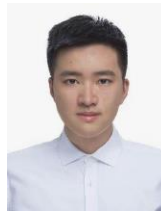

**Jintian Lin** received the B.E. degree and M.E. degree in Electrical Engineering from Southwest Jiaotong University, Chengdu, China in 2016 and 2019. And he is now working in Electric Power Research Institute, State Grid Zhejiang Electric Power Co.,Ltd.

His research is mainly focus on the analysis of frequency stability based on machine learning.

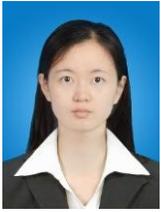

**Yichao Zhang** received the B.E degree in Electrical Engineering from the Southwest Jiaotong University, Chengdu, China, in 2017. And she is currently pursuing M.E degree in electrical engineering at Southwest Jiaotong University. She has been involved in research on the analysis of frequency stability based on the machine learning.

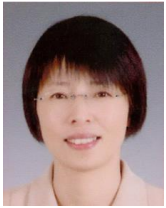

**Xiaoru Wang** received the B.E degree and M.E degrees in Electrical Engineering from Chongqing University, China, in1983 and 1988 respectively, and the Ph.D degree from Southwest Jiaotong University ,China, in 1998. She is currently a Professor at the School of Electrical Engineering of Southwest Jiaotong University. Her research interests include power system protection , stability analysis and control based on wide-area measurements. She is senior member of IEEE.

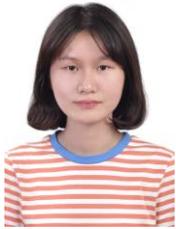

**Qingyue Chen** received the B.E. degree in Electric Engineering from Southwest Jiaotong University, Chengdu, China, in 2018. She is currently pursuing the M.E. degree in Electrical Engineering at Southwest Jiaotong University. Her current research interests include renewable energy generation and its frequency control.